
\nopagenumbers
\def\solar{\odot}
\def\gtorder{\mathrel{\raise.3ex\hbox{$>$}\mkern-14mu
             \lower0.6ex\hbox{$\sim$}}}
\def\undertext#1{$\underline{\smash{\hbox{#1}}}$}
\def\singlespace{\baselineskip=\normalbaselineskip}
\singlespace
\hsize=34pc
\vsize=52pc
\parindent=3pc
\overfullrule=0pt
\def\skip{\vskip12pt}
\def\nubare{\hbox{$\bar \nu_{\rm e}$}\tcar}
\def\tcar{\futurelet\next\testnextcar}
\def\testnextcar{\ifhmode\ifcat\next.\else\ \fi\fi}
\vglue1in
{\singlespace\indent
SUMMARY TALK:\ INTERNATIONAL SYMPOSIUM ON NEUTRINO\hfil\break
\indent ASTROPHYSICS, TAKAYAMA/KAMIOKA (10/'92)\smallskip}
\skip\skip\skip
John BAHCALL\hfil\break\indent
Institute for Advanced Study\hfil\break\indent
Princeton, New Jersey 0854\hfil\break\indent
USA
\skip\skip\skip
\noindent
\undertext{Contents}
\skip
1.\ The Conference

2.\ Solar Neutrinos

\ \ \ \ A.\ Operating Experiments

\ \ \ \ B.\ New Funded Experiments

\ \ \ \ C.\ Are Solar Neutrino Fluxes Time Dependent?

\ \ \ \ D.\ Have $pp$ Neutrinos Been Detected?

\ \ \ \ E.\ Is Directionality Required for Astronomical Observations?

\ \ \ \ F.\ Is There a Solar Neutrino Problem?

\ \ \ \ G.\ What Have We Learned?

3.\ Supernova Neutrinos

4.\ Atmospheric Neutrinos

5.\ High-Energy Neutrinos

6.\ Our Field Is Flourishing
\skip\noindent
1.\ \undertext{The Conference}
\skip
This has been a great conference, packed with important science and
arranged with style, tact, thoughtful consideration for the participants,
and deep
scientific insight.  It is one of the best organized
conferences I have attended.  The character of the organization
is typified by the quiet way a young student boarded our bus just before
we departed at 7:30 A.M. on our trip to the historic
Kamiokande facility.  The student asked that all of the speakers for
that day raise their hands and then went to each speaker to tactfully
inquire if they had remembered to bring their viewgraphs.  I am sure I
speak for all of us who attended the conference in expressing our
gratitude to J. Arafune (Chair of the Organizing Committee), to Y. Totsuka
(Co-Chair), and to K. Nakamura (Scientific Secretary), and to their many
younger associates.  All of these people worked hard
and effectively to maximize the
scientific benefit and the personal enjoyment for each participant.  I
 found it especially exciting to meet the excellent young Japanese
scientists whose names were known to me previously only from their
appearance on a long list of Kamiokande authors.

The international and local organizing committees managed to bring
together, in addition to the many younger people flushed with the excitement
of developing new ideas, the Four Founding Fathers of Neutrino
Astrophysics.  It was great to see Davis, Koshiba, Reines, and Zatsepin,
together with their (intellectual) sons, Kirsten and Totsuka,
struggling with the latest scientific challenges and providing
inspiration and insight to all of us.  In their epochal experimental
researches these individuals have demonstrated originality, perseverance
(i.e., stubbornness), and enormous achievement.  Their combined presence
enhanced the scientific success of the meeting.

In this summary talk, I will discuss four themes of the meeting:\ solar
neutrinos, atmospheric neutrinos, supernova neutrinos, and high energy
neutrinos.  There were other topics covered beautifully at the meeting,
including baryon number violation and grand unification, X-ray astronomy,
$\gamma$-ray astronomy, gravitational waves, dark matter, and cosmology.
 I will not cover these non-neutrino talks here in order to keep this
summary to a reasonable length.  For simplicity, I will not list
 references for material discussed in the talks,
but will instead give the name of the speakers at the
conference whose contributions in this book cover the topic under
discussion. The
statements based on the standard solar model are taken from
Bahcall and Pinsonneault (Rev. Mod. Phys., {\bf 64}, 885, 1992); this
paper
presents the only published solar eigen-models in which helium diffusion
was included in the calculations of the neutrino fluxes.  The
justification for other
statements made without explicit attribution about solar neutrinos or
about supernovae can be
found in {\it Neutrino Astrophysics}, Cambridge University Press (1989).
\skip\noindent
2.  \undertext{Solar Neutrinos}
\skip
Solar neutrinos was the most extensively discussed topic at the meeting;
it is the most thoroughly developed of the four areas of neutrino
research we discussed.
\skip\noindent
A.\ \undertext{Operating Experiments}
\skip
Table~1 summarizes the results presented at this conference for the four
solar neutrino experiments that are currently taking data, the $\rm
^{37}Cl$, Kamiokande ($\nu + e^-$ scattering), SAGE ($\rm ^{71}Ga$), and
GALLEX ($\rm ^{71}Ga$) experiments.  I list in Table~1 the energy
threshold, the neutrino sources, measured rate, and calculated rates
(Standard Solar Model).  The measured rates are all given at this
conference by Davis ($\rm ^{37}Cl$), Y. Suzuki (Kamiokande), Zatsepin
(SAGE), and Kirsten (GALLEX); the rates for the standard solar model are
from Bahcall and Pinsonneault.  I have
listed the average counting rates in Table~1.  Since the neutrino fluxes
are predicted to be constant to measurable accuracy
by the standard solar model, one must use
the average fluxes for tests of this model.  The last column contains
the maximum likelihood estimate for the total number of neutrino events
counted to date in each experiment.

I have given in the first column the names of the speakers who
described each of these experiments so that you can find the details in
the conference proceedings.

Of special note in Table~1 is the large difference, a factor of almost
four (many standard deviations), between the calculated
and the observed rates for the chlorine
experiment and the more moderate discrepancy, a factor of two, for the
neutrino-electron scattering experiment.  The
chlorine experiment is sensitive to neutrinos above 0.8~MeV whereas
Kamiokande is only sensitive to higher energy neutrinos (above 7.5~MeV).
If new physics is operating, the different energy sensitivities could
explain the
fact that the chlorine and the Kamiokande experiments differ from
the standard model predictions by two and four, respectively.

$$\vbox{\hsize=14.2cm\halign{#\hfil\tabskip=.5em plus.5em
minus.5em&\hfil#&\hfil#\hfil&\hfil#\hfil&\hfil#\hfil
&\hfil#\hfil\tabskip=0pt\cr
\multispan6{Table 1.\ Solar Neutrino Experiments.\hfill}\cr
\noalign{\medskip\hrule\medskip}
\hfil Target\hfil&Energy&Source(s)&Measured&SSM$^\dagger$&Neutrino\cr
&(MeV)&&&&Counts\cr
\noalign{\medskip\hrule\medskip}
${\rm ^{37}Cl}$&$\geq 0.8$&All but $pp$&$2.23 \pm 0.23$&$8.0 \pm
1.0$&750\cr
(Davis)\cr
\noalign{\bigskip\smallskip}
$\nu + e$&$\geq 7.5$&${\rm ^8B}$&$0.49 \pm 0.04 \pm 0.06$&1.0&380\cr
(Suzuki)&$\geq 18$&$hep$&$< 49$&1.0\cr
&13&$\bar\nu_e$&$ < 0.06\ \phi_{\rm SSM}({\rm ^8B})$&0.0\cr
\noalign{\bigskip\smallskip}
${\rm ^{71}Ga}$&$\geq 0.2$&All&$58^{+17}_{-24} \pm
14$&$132^{+7}_{-6}$&25\cr
(SAGE, Zatsepin)\cr
\noalign{\bigskip\smallskip}
${\rm ^{71}Ga}$&$\geq 0.2$&All&$83^{+20}_{-19} \pm
8$&$132^{+7}_{-6}$&75\cr
(GALLEX, Kirsten)\cr
\noalign{\medskip\hrule\medskip}
\noalign{\vbox{\hsize=14.2cm \parindent=10pt
\item{$^\dagger$}Average over time. Statistical errors
quoted are effective
1$\sigma$ values; statistical errors precede systematic errors.}}\cr
}}
$$

The two gallium experiments, GALLEX AND SAGE,
are in satisfactory agreement with each
other, although with their currently
large statistical errors it is difficult to
say anything very definitive until they have been operating for a
longer period.
The gallium calibration measurements with intense $^{51}$Cr sources
will provide very important tests of our understanding of how these
experiments work; the calibration measurements will begin in about a year
for both experiments.
Ultimately, when the two independent experiments have been operating for
several years---and when both have been calibrated with external
radioactive sources---we will have have a very reliable number for the solar
neutrino capture rate on gallium.
\skip\noindent
B.\ \undertext{New Funded Experiments}
\skip
Table~2 describes the four new solar neutrino experiments that have been
funded for development.  Each of the modes of each of the experiments
listed in Table~2 is expected to yield more than 3,000 neutrino events
per year (except for the $\nu - e$ scattering mode of SNO).
In one year, each experiment will record
three or more times the total number of
neutrino events that have been counted to date in all solar neutrino
experiments since the chlorine experiment began operating a quarter of a
century ago.  With this much greater statistical accuracy, we can
expect our subject to become much more precise.

The experiments are listed in order of their expected completion
 dates, SNO (1995), Super Kamiokande (1996), BOREXINO
($\sim$~1996), and ICARUS (1998).  Table~2 lists the neutrino
threshold energy for
each reaction mode and the individual reactions that will be observed.

$$
\vbox{\halign{#\hfil\tabskip=2em plus2em
minus2em&\hfil#\hfil&\hfill$#$\tabskip=0pt&$\ #$\hfill\cr
\multispan3{Table 2.\ \vtop{\hbox{New Solar Neutrino
Observatories.}\hbox{Typical Event Rates $\gtorder~3 \times
10^3$~yr$^{-1}$}}\hfill}\cr
\noalign{\medskip\hrule\medskip}
\hfil Observatory\hfil&$E_{Th}(\nu)$
&\multispan2{\hfil\rm Reaction(s)\hfil}\cr
&(MeV)\cr
\noalign{\medskip\hrule\medskip}
SNO&6.4&\nu_e + {\rm ^2H} \to& p + p + e^-\cr
(Ewan)&2.2&\nu + {\rm ^2H} \to& n + p + \nu\cr
&5&\nu + e^- \to& \nu + e^-\cr
\noalign{\bigskip\smallskip}
Super-&5&\nu + e^- \to& \nu + e^-\cr
Kamiokande\cr
(Takita)\cr
\noalign{\bigskip\smallskip}
BOREXINO&0.25&\nu (^7{\rm Be})+ e^- \to& \nu (^7{\rm Be}) + e^-\cr
(Raghavan)&&\cr
\noalign{\bigskip\smallskip}
ICARUS&$\sim 10$&\nu_e + {\rm ^{40}Ar} \to& e^- + {\rm ^{40}Kr}^*\cr
(Revol)&5&\nu + e^- \to& \nu + e^-\cr
\noalign{\medskip\hrule}
}}
$$

The SNO experiment has, as George Ewan told us, two unique
capabilities:\ SNO will measure precisely the energy
spectrum of electron-flavor neutrinos
above about 5~MeV in the charged current reaction (neutrino absorption
by deuterium)
and will measure the total neutrino flux independent of flavor in the
neutral current reaction (neutrino disintegration of deuterium). The
comparison of the fluxes measured via neutrino absorption on deuterium
and by neutrino disintegration of deuterium will
test the equality of the charged and the neutral currents.
If the total neutrino flux is not equal to the electron neutrino flux,
then we will have direct evidence for neutrino flavor changing. (The two
measured fluxes would also be equal if electron neutrinos change only into
sterile neutrinos.)

I am
delighted that the SNO collaboration is giving first priority to
determining the shape of the $\nu_e$ energy spectrum; the shape of this
function is independent of solar astronomy (to an accuracy of one part
in $10^5$, Phys. Rev. D, {\bf 44}, 1644, 1991).  A measurement of the
neutrino energy spectrum could establish that physics beyond the
standard electroweak model is required, independent of any consideration
of solar physics.
The charged and neutral currents must be
equal unless some neutrinos change their flavor after they are created
in the solar core.  Unfortunately, no energy information will be
available for the neutral current detection.  Also, the neutral and
charged-current fluxes would be equal even if some of the original
electron-type neutrinos changed into sterile neutrinos.

I was excited to hear from Revol
that the ICARUS collaboration has successfully completed testing a 3-ton
prototype.  The welcome news about ICARUS indicates that this very important
experiment, which will have a unique ``smoking-gun'' mode of detection of solar
neutrinos (gamma rays accompanying neutrino-produced electrons), will
proceed as planned.

Note that some redundancy will be available if all four experiments
operate as planned.  The ICARUS experiment will provide a precise
measurement of the $\nu_e$ energy spectrum above about 10~MeV, which can be
compared with the energy spectrum measured by SNO.  Three experiments
(Super Kamiokande, SNO, and ICARUS) will measure for ${\rm ^8B}$
neutrinos the $\nu_e - e$
scattering rate and the recoil electron energy spectrum; the electron
recoil spectrum reflects
the incoming neutrino energy spectrum.  Takita showed us, for example,
how well Super Kamiokande can distinguish---by measuring
the electron recoil energy spectrum---between the small difference in
the electron recoil spectrum for the two MSW
solutions that are allowed by the existing experiments.

The BOREXINO experiment is absolutely essential.  It is the only
experiment planned for this decade that will measure the energy of
individual events with energies less than 5~MeV.  This information is
necessary in order to determine the neutrino survival probability at low
energies. The BOREXINO experiment also
has another highly desirable feature; it measures the $\nu_e$ flux at a
specific energy.  One does not have to average the theoretical
predictions for BOREXINO over energy (and therefore
lose uniqueness) as is the case
for electron scattering by a continuum neutrino energy source.

The most urgent need in this field is for other low-energy neutrino
detectors.  No practical solution appears in sight for the fundamental
problem of detecting the basic $pp$-neutrinos and only one funded
experiment will observe the individual $\rm ^7Be$ neutrino reactions.
Kopylov described a possible $\rm ^7Li$ radiochemical experiment, but a
method still must be developed for counting the $\rm ^7Be$ atoms that
are produced.

Two $\nu_\mu - \nu_\tau$ oscillation experiments at CERN, called CHORUS
and NOMAD, will begin operating soon. Although they are vacuum
oscillation experiments, CHORUS and NOMAD have the sensitivity in mass
to test some of
the MSW solutions of the solar neutrino problem that were discussed at
this conference.  There were no talks on these important experiments,
but I think I remember reading that they are expected to be sensitive to
mass differences for $\nu_\mu - \nu_T$ of the general order of
$\gtorder 10\ {\rm eV}^2$ with $\sin^2 2\Theta \gtorder
 10^{-3.5}$ or so.  A similar experiment, NUMI, is planned at FNAL.  These
laboratory experiments are important tests of standard model physics and
may, in addition, be crucially related to the astronomical
interpretation of the observed solar neutrino fluxes.
\skip\noindent
C.\ \undertext{Are Solar Neutrino Fluxes Time Dependent?}
\skip
Do the observed event rates vary with time?  They are not expected to do
so since the characteristic cooling time in the solar core is of order
$10^6$~years and since the important nuclear reaction rates have
lifetimes between $10^4$ and $10^{10}$~years.  The only way that has
been found to account for the suggested $\sim~11$~year cyclic dependence
is to invoke new physics incompatible with the standard electroweak
model, a large neutrino magnetic moment (discussed at this conference by
Lim).

The Kamiokande data are not correlated with sunspot intensity, as was
shown clearly by Y. Suzuki in his lecture.  Davis pointed out that the
chlorine data also show no indication of a correlation with sunspots
during the period in which the Kamiokande experiment was operating.  What's
more, the chlorine data do not indicate any correlation for the first
seven year period of data taken between 1970 and 1977.
The data do show an apparent
dependence on sunspot cycle for the two cycles that are sandwiched
between apparently constant rates.  My view is that this seeming
correlation is most likely a fluke occurrence in noisy data, but it is
very important to continue the chlorine measurements
to obtain information about any
long-term trends or time-dependences in the data.
This is especially true since, as Kopylov told us,
the chlorine experiment does not appear likely to be repeated.

Fortunately, the average rate observed in the chlorine experiment is
robust.  The average rate observed in the initial (non-varying) period
from 1970--1976 is $(2.0 \pm 0.35)$~SNU; the
average rate observed (when variations
have been suggested) from 1977--1990 is $(2.1 \pm 0.2)$~SNU and,
as Davis told us in
his talk, the average of all the data is 2.23~SNU.  I conclude that the
average rate of neutrino captures observed over a long period (a number
of years) is well determined and it is this quantity that we need
to test the standard model.

\skip\noindent
D.\ \undertext{Have $pp$ Neutrinos Been Detected?}
\skip
What do the gallium experiments tell us about the fundamental $pp$
neutrinos?  Radiochemical experiments have only an energy threshold but
no energy resolution.  Therefore, all that we know is that, e.g., for the
GALLEX experiment the observed event rate is approximately $83 \times
10^{-36}$ captures per target atom per sec, i.e., 83~SNU.  To go
further, we must compare the observed event rate with the expected event
rates from different sources as predicted by the standard solar model.
The total predicted rate from all solar sources is
132~SNU and the rate predicted from $pp$
(and $pep$) neutrinos alone is 74~SNU.  The similarity between the $pp$
rate in the standard solar model, 74~SNU, and the observed GALLEX rate,
83~SNU, has suggested to some that GALLEX has ``definitely" observed the
$pp$ neutrinos.  But, the rate expected in the
standard solar model from all neutrinos except $pp$ neutrinos is also
reasonably close (with the current large statistical errors) to the
observed rate.  The ``all but $pp$ rate" is equal to:\  \hbox{36 (from ${\rm
^7Be}$)~+~8 (from CNO)~+~7} (from $\rm ^8B$; rate halved because of
Kamiokande result).  The total from ``all but $pp$" is therefore 51~SNU.

Table~3 compares the observed versus expected (standard solar model)
rates for the $pp$ neutrinos, the ``all but $pp$ neutrinos," and the
total standard model fluxes.  Both the GALLEX and the SAGE experimental
rates are within $1\sigma$ of the value expected for $pp$ neutrinos
only, but neither experiment is as much as $2\sigma$ away from the value
expected for ``all but $pp$."  Both experiments are more than
$2\sigma$ away from the standard solar model value (SAGE is more than $4\sigma$
away).  I conclude that there is some experimental indication that $pp$
neutrinos are being observed in the gallium experiments, but (and this
is usually not emphasized) the evidence for a discrepancy with the
standard solar model is stronger than the evidence that $pp$ neutrinos
are being detected.

$$\vbox{\halign{#\hfil\tabskip=2em plus2em minus2em
&\hfil#\hfil&\hfil#\hfil&\hfil#\hfil\tabskip=0pt\cr
\multispan4{Table 3.\ Observed Versus Expected Ga Rates.$^\dagger$\hfill}\cr
\noalign{\medskip\hrule\medskip}
\hfil Exp.\hfil&$|$Obs.--$pp$$|$&$|$Obs.--(all but $pp$)$|$&$|$Obs.--SSM$|$\cr
\noalign{\medskip\hrule\medskip}
GALLEX&9&32&49\cr
83 SNU&(0.5$\sigma$)&(1.7$\sigma$)&(2.5$\sigma$)\cr
\noalign{\medskip}
SAGE&16&7&74\cr
58 SNU&(0.9$\sigma$)&(0.4$\sigma$)&(4.3$\sigma$)\cr
\noalign{\medskip\hrule\medskip}
\multispan4{$^\dagger$\ Only statistical errors included in
$\sigma$.\hfill}\cr
}}
$$

What do we need to do to sharpen the test for the presence of $pp$
neutrinos?  The statistical errors must be reduced in the
gallium experiments.  But, it is also essential to determine if the
gallium experiments are really observing 36~SNU from the $\rm ^7Be$
neutrinos, as suggested by the standard solar model.  Fortunately, as
Raghavan told us, BOREXINO should give us a good value for the $\rm
^7Be$ neutrino flux.
\skip\noindent
E.\ \undertext{Is Directionality Required for Astronomical
Observations?}
\skip
Suzuki showed a dramatic figure demonstrating that the neutrinos
observed by Kamiokande come from the direction of the sun.  This result
is of great importance scientifically and I personally was thrilled to
see this clear demonstration of solar origin (for amusement see the
first theoretical discussion of this effect in Phys. Rev. B, {\bf 135},
137, 1964).

I do not agree with the statement made by Koshiba that a solar neutrino
experiment must have directionality in order to be considered astronomy.
 The sun is the overwhelming source of locally-observed
neutrinos for the same reason that the sky is bright
in the day and is dark
at night.  A numerical calculation summarized in Section~6.5 of {\it
Neutrino Astrophysics} gives a ratio of more than $10^9$ for solar to
non-solar neutrinos in the Kamiokande energy range.  Moreover, the solar
neutrino flux is certainly less than or equal to the total
neutrino flux observed
in the (non-directional) chlorine and gallium experiments.
Since the total observed flux is less than the calculated
flux from solar models, the chlorine and gallium experiments provide
important information about the sun that does not depend upon
knowing that the sky is dark at night.
Of course, the most straightforward astronomical inferences assume---for
directional as well as non-directional experiments---that nothing
happens to the neutrinos
after they are created.

The last two Decade Surveys for Astronomy (conducted by the National
Academy of Sciences) in the United States have pointed to the chlorine
experiment as a major contribution to astronomical knowledge and have
strongly advocated the gallium experiments.  Wearing my hat as president
emeritus of the American Astronomical Society, I can say that we
astronomers are proud to claim the chlorine, gallium and the (future)
BOREXINO experiments as important astronomical observatories.
\skip\noindent
F.\ \undertext{Is There a Solar Neutrino Problem?}
\skip
Finally, we need to consider the fundamental question of our subject:
Is there a solar neutrino problem?
The answer is ``yes,'' because the chlorine and
Kamiokande experiments are inconsistent with the most accurately
calculated solar models.
In the 30 years that I have been calculating solar neutrino fluxes, no
one has produced a solar model with conventional physics that has
neutrino fluxes in agreement with the chlorine experiment.
The $\rm ^8B$ neutrino flux observed by
Kamiokande exceeds by $2\sigma$---if nothing changes the neutrino energy
spectrum--the equivalent event rate in the chlorine experiment.
Moreover, the two gallium experiments are more
than $2\sigma$ away from the standard model predictions.

The most decisive way of exhibiting the discrepancy between the standard
model calculations and the observations is to show the results of the
Monte Carlo calculation for 1000 solar models when the \b8 fluxes
for each model are replaced with the experimental value obtained by the
Kamiokande experiment.  This histogram is shown in Figure~1, taken from
a recent paper by Hans Bethe and myself.
Even if the uncertainties associated with the
calculation of the \b8 flux are artificially removed by adopting the
Kamiokande value for this quantity, the calculated event rate is
inconsistent with the observed event rate in the chlorine experiment.
Not one model in 1000, even with the measured Kamiokande flux, is within
the $3 \sigma$ error limit of the chlorine experiment.  In my view, this is
the core of the ``solar neutrino problem.''

It is of course possible that one or more of the existing experiments is
incorrect.  I do not think this is likely, but the history of science
has taught us that important experiments must be confirmed by different
techniques in order to derive secure conclusions. Any new experiment
that can provide unambiguous information about solar neutrino fluxes
is highly desirable.

\skip\noindent
G.\ \undertext{What Have We Learned?}
\skip
The Kamiokande and chlorine experiments show that---although there is a
discrepancy with the standard solar model---the ${\rm ^8B}$ neutrino
flux at the earth is $\sim 3 \times 10^6~{\rm cm}^{-2}{\rm s}^{-1}$.
This point was stressed by Ray Davis.

{}From an astronomical point of view, this is a great achievement.  The
fact that the standard solar model predicts a ${\rm ^8B}$ neutrino flux
within a factor of two of the measured value is a unique and sensitive
confirmation of the basic ideas of stellar evolution and of nuclear
energy generation in stars.

If all four of the operating experiments
are correct, then the chlorine and Kamiokande results show that, in
addition, new
physics is required (see Figure~1).  The gallium experiments then fix
(approximately) the MSW parameters that can describe all the observational
results.  As Smirnov and Pakvasa told us, the standard MSW solutions
are then: $\delta m^2~\sim~10^{-6}~{\rm eV}^2 (\sin^2~2\Theta \ll 1)$ or
$\delta m^2~\sim~10^{-5}~{\rm eV}^2~{\rm to}~10^{-4}~{\rm eV}^2
(\sin^2~2\Theta \sim 0.5)$.  There is an additional allowed region with
$\delta m^2 \sim 10^{-6}~{\rm eV}^2~{\rm to}~10^{-8}~{\rm eV}^2$, also at
large mixing angles ($\sin^2~2\Theta \sim 0.5$).

The new experiments listed in Table~2 will
either confirm and extend these conclusions or reveal that there is some
(currently unknown) fundamental error in one of the experiments or in
the solar model calculations.

\skip\noindent
3.\ \undertext{Supernova Neutrinos}
\skip
We heard clear descriptions of the theoretical expectations for
supernova neutrinos and, with great honesty, the difficulties that must
be overcome to make accurate calculations.  The talks by Sato, Janka,
H. Suzuki, Burrows, and Nomoto convinced me that the problems are
fascinating as well as formidable.  It is clear that the theory is on
the right track, since the pre-existing predictions gave for
supernova 1987A the total neutrino energy correct to better than an order of
magnitude, and also gave good estimates for the average
temperature of the electron anti-neutrinos and for the duration of the
burst.  These later two calculations are particularly impressive since
they depend upon the result of neutrino trapping as the neutrinos diffuse
out of the stellar core, which converts a
population of several hundred MeV neutrinos to a swarm of neutrinos with
characteristic energies of order 10~MeV or less.

Each time that I get a little depressed by the seemingly endless detail
and the great precision that are required to compute an accurate solar
model, I think of the much greater difficulties that my friends working
in the field of supernova neutrinos must overcome.  As a measure of the
relative difficulty of the solar and supernova problems, and as an
indication of how much smarter the supernova theorists have to be than I
am, here are some relevant dimensionless ratios:
$$
{\rho ({\rm neutron\ star}) \over \rho (\solar)}~\sim~10^{+12}~,~{\tau
(\solar) \over \tau ({\rm supernova})}~>~10^{18}~,~{E ({\rm
supernova})\over E ({\rm solar\ flare})}~>~10^{20}~.
$$
Here $\rho$ is the local density, $\tau$ is the lifetime (of the main
sequence sun or the supernova collapse), and $E$ is the total energy (of
the supernova explosion or of a large solar flare).

We heard descriptions of several existing and potential supernova
detectors at this conference, including Kamiokande (Y. Suzuki), MACRO
(Spinetti), Baksan (Kopylov), SNO (Ewan), Super Kamiokande (Takita),
ICARUS (Revol), Soudan 2, and LVD.  Once all of these
detectors are operating, we
will be well equipped to study Galactic supernovae.  In fact, the
Kamiokande, MACRO, Baksan, and LVD detectors are already obtaining
valuable limits on the possible occurrence of supernovae anywhere in the
Galaxy (including regions that are hidden by dust from visible
observations).

Krauss presented some results of a menu-driven
Monte Carlo simulation program that illustrate what can be learned
about supernovae and about neutrino physics by observing an explosion
with Kamiokande or Super Kamiokande.  The results were very interesting
and in some cases unexpected.  The difficulty of detecting the initial
neutronization burst, for example, was surprising but in retrospect
understandable.

Koshiba proposed in his talk that a very large detector be built to
observe extragalactic supernovae and Cline proposed a network of
supernovae detectors.  Both of these proposals are aimed at achievable
and fundamental astronomy and physics goals.  Either of the proposals is
feasible, in my opinion, provided a group of people dedicate themselves
to making the dreams a reality over the approximately one decade
necessary to persuade the community, to obtain funding, and to construct the
detectors.

Tammann described a determination of the rate for Galactic supernovae of
one per $33^{+12}_{-7}$ years, using both extragalactic and Galactic
supernovae.  I found Tammann's description of his analysis fascinating,
including the difficult factors that must be understood such as the
effects of dust hiding Galactic supernovae, the small number (5) of
historic Galactic supernova, and the effects of galactic type,
inclination, and $H_0$ in the determination of the rates of
extragalactic supernovae.
One of the crucial elements in Tammann's determination is the
calibration of the efficiency of the amateur astronomers in detecting
supernovae in extragalactic nebulae.  I could not erase from my mind the
irreverent image of Tammann on a dark night creeping up to Reverend Evans as he
concentrated on the heavens and measuring the dilation of the Reverend's
pupils with a micrometer.

A number of different methods have been applied to the calculation of the
Galactic supernova rate and these yield expected
rates ranging from the lowest calculated value of 11
years (from the computed death rate of massive stars) to more than 60
years (from pulsar statistics).
As Tammann told me privately, ``the
determination of the Galactic supernovae rate is harder than measuring
$H_0$," for which he is also famous.
The systematic errors in the subject should be reduced when the results
of the Berkeley automatic supernova detection survey and similar surveys
have accumulated sufficient observing time.

\skip\noindent
4.\ \undertext{Atmospheric Neutrinos}
\skip
The ratio of muon to electron neutrinos from atmospheric cascades was
reported to be lower than expected in both the
Kamiokande (Kajita) and the IMB (Dye)
experiments.  In both cases, the observed ratio, $R$, of $\nu_{\mu}$ to
$\nu_{e}$ events is less than
the ratio expected from Monte Carlo simulations.
No matter how the experimental cuts were made, the ratio of ratios was
always found to be approximately in the range:

$$ {{R\left(\mu /e \right)}_{\rm Obs.} \over
{R\left(\mu /e \right)}_{\rm Monte Carlo}}
{}~=~ \left(0.60 ~{\rm to} ~ 0.65 \right)~ \pm 0.05 .
$$
This result is highly significant statistically, but the systematic
errors require further investigation.

Gaisser and Honda gave explicit discussions of the ingredients that
went into the predictions of the $\nu_{\mu}$ to
$\nu_{e}$ ratio and stressed the various
uncertainties in the input data of the theoretical calculations.
The robustness
of the theoretical predictions is remarkable.
Gaisser stressed that this is partly because the uncertainties in the
primary cosmic ray beam that produce all of the secondaries affect both
types of neutrinos in the same way.  Also, the uncertainties in the data
on pion and kaon production as the cosmic rays cascade through the
atmosphere affect both types of neutrinos similarly.

I was puzzled by the fact
that many of the senior experimentalists with whom I discussed the
problem---some coauthors on the papers being presented---felt that the
discrepancy might ``go away," but could not highlight specific
weaknesses in the arguments.  Perhaps they were influenced by the 15\%
change in the calculated ratio due to muon polarization (pointed out by
Volkova) and they are waiting for another couple such revisions.

The measurements of the $\nu_\mu/\nu_e$ ratio
could have wide-ranging consequences for physics,
including the possibility of a $\nu_{\mu}$ to
$\nu_{e}$ mixing with a mass difference $\delta m^2 ~\gtorder~ 10^{-2}$
${\rm eV^2}$.  However, I think we need to understand in more
detail all aspects of the problem.
Most importantly, we need to identify clearly
and scrutinize most carefully those aspects of the
calculation that can affect the predicted $\nu_\mu/\nu_e$
ratio such as:
the experimental
discrimination between $\nu_{\mu}$ and $\nu_{e}$ events,
the guts of the Monte Carlo simulations, the
implications of the discrepancies between the calculated and observed
absolute rates.
the percentage of produced muons that decay in transit;
the composition of the primary cosmic rays (which affects the
ratio of $\nu_e$ to \nubare), and
the relevant cross sections for quasi-elastic neutrino interactions in
nuclei.
Precise measurements of the primary flux and of the muon flux at high altitudes
would be helpful in resolving some of the ambiguities in the neutrino flux
calculations, as stressed by Gaisser.

In the early days of the solar neutrino problem, there were several
intense meetings with small groups in which Ray Davis and I described our
respective parts of the problem and experts in chemistry, in nuclear
physics, in high energy physics, and in astronomy
probed for the weak spots in the arguments.  These meetings were
informal study sessions, much like oral Ph.D. examinations, where
questions could be examined and discussed in depth.  They had the
opposite flavor from the currently-popular large conferences.
A number of tests and checks, both experimental and theoretical,
were suggested in these small meetings that
led all of us to understand the problem better and to ultimately have
confidence in what we are doing.  Perhaps something similar should be
tried with the atmospheric $\nu_{\mu}$ to
$\nu_{e}$ ratio.  A group of
people interested in the details and expert in different aspects of the
subject could get together with the
experimentalists and theorists involved in the problem for some detailed
but essentially private discussions of what is going on.  My expectation
is
that this would lead to a better understanding and ultimately a greater
confidence in the final results.
\skip\noindent
5.\ \undertext{High-Energy Neutrinos}
\skip
The possibility of observing high-energy neutrinos ($\gtorder 100$ GeV) with
AMANDA (Halzen) or DUMAND (Learned) is thrilling.  There are fabulous
possibilities for detecting different kinds of
sources,including the debris from high-energy cosmic
ray interactions
in the interstellar medium, dark matter annihilation in the sun, and
shock waves in active galactic nuclei.  In all cases, the basic process
is protons crashing into something and thereby producing pions and muons
that decay into neutrinos (and other less penetrating products).

Halzen pointed out that if the high-energy gamma rays from Mrk~421 are
produced by proton collisions instead of by the inverse
Compton effect then AMANDA should see many events per
year from this source.  (An angular resolution of order one degree is
expected.)  In such a case, there could well be many
more high-energy neutrino sources than high-energy gamma ray sources
since intergalactic starlight probably creates an appreciable opacity to
high-energy gamma rays but not to high-energy neutrinos.

AMANDA and DUMAND will explore a frontier of the Universe.  It is hard
to know what they will find, but the results might well dominate
meetings on Neutrino Astrophysics in the second half of this decade.
\skip\noindent
6.\ \undertext{Our Field Is Flourishing}
\skip
Solar neutrino astronomy has become a mature
field with four operating experiments and four more planned experiments
that will increase the data rate by almost two orders of magnitude.
Until the new experiments are operating, we will
have to make use of solar model calculations of the expected fluxes
in order to interpret the results.
Doing this, we mix together possible new physics with possible new astronomy.
One of the most wonderful aspects of
the new experiments is that they will make possible precision
tests---independent of solar considerations---of the
shape of the $^8$B energy spectrum and of the ratio of charged to
neutral current fluxes.
In addition, we should be able to see for the first time individual
events from the low-energy $^7$Be neutrino line.  The results of the
new experiments will be rich in the information they provide about
physics and about astronomy.
At present, we can say that all four of the existing
solar neutrino experiments disagree with the standard solar model and
that the chlorine and Kamiokande experiments cannot both be correct {\it
if the $^8$B solar neutrino spectrum is not changed by new physics}.
I believe that it is likely that new physics is being revealed by
the existing
solar neutrino experiments, but history teaches us that
we must be cautious about reaching conclusions until the
results are confirmed by different experiments.  It is possible that the
new CERN and FNAL $\nu_\mu - \nu_\tau$ oscillation experiments will show
that neutrino oscillations---the most attractive theoretical explanation
of the solar neutrino problem---do occur in nature.

Several neutrino detectors are currently
operating that could provide detailed information about Galactic supernova
explosions and perhaps even reveal a cosmologically interesting neutrino
mass.  As we gain more experience
with the existing and under-construction new detectors, I hope  we will be
encouraged to build even larger detectors capable of reaching to
distances sufficient to detect every year neutrinos from
supernova explosions.

We can look forward to a more intense
scrutiny of the $\nu_\mu$ to $\nu_e$ ratio in the atmospheric cosmic
rays.  New experiments (such as MACRO), and more detailed examination of the
already-available results, should clarify this puzzle and perhaps
present us with another well-defined problem that could conceivably
point to physics beyond the standard electroweak model.

Finally, there
is the awesome possibility that high-energy neutrinos will be observed
in AMANDA, DUMAND, and other large-area detectors,
extending the frontier of astronomy to high-energy
neutrinos and opening a new window to the Universe.
\end